\documentclass[11pt,a4paper]{article}
\usepackage{float}
\usepackage{multirow}
\usepackage{graphics}
\usepackage[cp1252,utf8]{inputenc}
\usepackage{hyperref}
\usepackage{graphicx} 
\usepackage{amsmath}
\usepackage{amsfonts}
\usepackage{amssymb}
\usepackage{bigints} 
\usepackage{relsize} 
\usepackage[top=1in, bottom=2cm, left=2cm, right=2cm]{geometry}
\usepackage{authblk}

\usepackage{changepage} 

\usepackage{afterpage}

\usepackage{placeins}

\title{\bf Entropy function from the Einstein boundary term}
\author[a]{\bf  Anirban Roy Chowdhury \thanks{iamanirban@bose.res.in}}
\author[b]{\bf  Ashis Saha \thanks{ashisphys18@klyuniv.ac.in}}
\author[c]{\bf Sunandan Gangopadhyay \thanks{sunandan.gangopadhyay@bose.res.in}}

\affil[a,c]{\textit{Department of Theoretical Sciences,
 S.N.~Bose National Centre for Basic Sciences,}
\textit{JD Block, Sector-III, Salt Lake, Kolkata 700106, India}}
\affil[b]{\textit{Department of Physics, University of Kalyani, Kalyani 741235, India}}

\date{}

\begin{document}

\maketitle

\begin{abstract}

\noindent We show using the entropy function formalism developed by Sen \cite{Sen:2005wa} that the boundary term which arises from the Einstein-Hilbert action is sufficient to yield the Bekenstein-Hawking entropy of a static extremal black hole which is asymptotically flat. However, for asymptotically $AdS$ black holes, the bulk term also plays an important role due to the presence of the cosmological constant. Further, we show that for extremal rotating black holes, both the boundary and the bulk terms contribute non-vanishing pieces to the entropy.

\end{abstract}


\section{Introduction}
Classically black holes are the solutions of the Einstein field equations with the special property that they have a hypothetical surface known as the event horizon which acts as a one way membrane , and these are objects with absolute zero temperature. However, according to quantum theory, black holes behave as perfect black bodies with finite temperature known as the Hawking temperature ($T_H$), and as a consequence it emits Hawking radiation in accordance with the laws of black hole thermodynamics \cite{Hawking:1974rv,Hawking:1974sw}. As a thermodynamic system, the thermal entropy of a black hole, known as the Bekenstein-Hawking entropy , is proportional to the area of its event horizon \cite{Bekenstein:1973ur}. This reads 
\begin{equation}
S_{BH}=\frac{A_{EH}}{4G}~
\end{equation}
where $A_{EH}$ is the area of the event horizon of the black hole and $G$ is the Newton's gravitational constant.

\noindent Interestingly, black holes can have entropy even at zero temperature, and black holes which have zero temperature are said to be extremal black holes. A powerful method to calculate the entropy of extremal black holes (using the near horizon geometry) is the entropy function formalism developed by Ashoke Sen in \cite{Sen:2007qy}. Since then there has been a lot of research in this area
In this regard we would like to mention that the Noether charge approach to determine the entropy of a black hole for non-extremal black hole does not work for an extremal black hole. This is because the relation between the charge $Q$ and the entropy $S$ of a black hole is of the form $Q=T_{H}S$ \cite{Wald:1993nt,Iyer:1994ys,Majhi:2012tf,Majhi:2012nq}. But for an extremal black hole $T_{H}=0$ and hence $S$ cannot be determined from this relation. Hence the entropy function formalism finds its importance in determining the entropy of extremal black holes.
The entropy function approach introduces a function which is an integration over the transverse coordinates of the metric of the Lagrangian density of the Einstein-Hilbert action and the matter sector. The spacetime metric used in this definition of the function is the near horizon geometry of the black hole. It turns out that for spherically symmetric black holes in $3+1$-dimensions, the near horizon geometry has the form $AdS_{2}\times S^{2}$, and for rotating extremal black holes the near horizon geometry has the form $AdS_{2}\times U(1)$. This observation is independent of any coordinate system which one can possibly use to write down the metric. Hence, an ansatz of the near horizon geometry is taken to be of this form with undetermined quantities in it. The formalism determines these quantities and hence given the near horizon geometry. A new function is then defined in terms of the electric field strength. This function is said to be the entropy function. It turns out that this function computed using the parameters which extremizes the function yeilds the entropy of the extremal black hole. \\
In this paper we show that the entropy function formalism can be developed from the Einstein boudary term in the presence of the matter sector of the theory. This is because the bulk term in the Einstein-Hilbert action do not contribute to the entropy function. Hence it appears that the surface term plays a crucial role in constructing the entropy function. However, we shall show that this observation is not true for extremal static asymptotically $AdS$ black holes and extremal rotating black hole solutions. In both these cases, the bulk term provides a non-vanishing contribution to the entropy function. To establish these observations, we discuss the entropy function formalism for three different black holes, namely, the extremal Reissner-Nordstr\"om black hole, the extremal charged BTZ black hole, and the extremal Kerr black hole.

The paper is organized as follows. In section (\ref{sec1}), we take a brief look into the  Einstein boundary term from the Einstein-Hilbert action. A discussion on the nearhorizon geometry of an extremal black hole spacetime is given in section (\ref{cc}). In section (\ref{aa}), we develop the entropy function formalism using the Einstein boundary term. Here we consider three extremal black holes, namely, the extremal  Reissner-Nordstr\"om black hole, the extremal charged BTZ black hole, and the extremal Kerr black hole.

\section{A brief look at the Einstein boundary term}\label{sec1}
{We start our analysis with the Einstein-Hilbert (EH) action with a cosmological constant together with the GHY boundary term \cite{York:1972sj,Gibbons:1976ue,York:1986lje}
\begin{equation}\label{1}
\mathcal{A}=\frac{1}{16\pi G}\int d^{4}x\sqrt{-g} \left(R-2 \Lambda \right)+\frac{1}{8\pi G}\int d^{4}x \partial_{a}(\sqrt{-g}K N^{a})~\equiv~\mathcal{A}_{EH}+\mathcal{A}_{GHY}~
\end{equation}
where $R$ is the Ricci scalar and $\Lambda$ is the cosmological constant. The importance of the GHY boundary term lies in the fact that it leads to a well defined boundary value problem. 
A simple calculation shows that one can split $\mathcal{A}_{EH}$ into a bulk and a surface term . This reads \cite{Einstein:1916cd}
\begin{eqnarray}\label{2}
\mathcal{A}_{EH}&=&\frac{1}{16\pi G}\bigg[\int d^{4}x\sqrt{-g}~M_{a}{^b}{^c}{^d}R^{a}{_b}{_c}{_d}-\int d^{4}x ~2\Lambda \bigg]\nonumber\\
&=&\frac{1}{16\pi G}\int d^{4}x\sqrt{-g}\big(\mathcal{L}_{quad}-2\Lambda\big) + \frac{1}{16\pi G}\int d^{4}x\mathcal{L}_{surf} 
\end{eqnarray}
\noindent where $ \mathcal{L}_{quad}$ , $\mathcal{L}_{surf}$ and $M_{a}{^b}{^c}{^d}$ are given by 
\begin{eqnarray}
\mathcal{L}_{quad}&=&2 M_{a}{^b}{^c}{^d}\Gamma^{a}_{dk}\Gamma^{k}_{bc}\nonumber\\  \mathcal{L}_{surf}&=&2\partial_{c}\big[\sqrt{-g}M_{a}{^b}{^c}{^d}\Gamma^{a}_{bd}\big]\nonumber\\
M_{a}{^b}{^c}{^d}&=&\frac{1}{2}(\delta^{c}_{a}g^{bd}-\delta^{d}_{a}g^{bc})~.
\end{eqnarray}
The first term is a bulk term and the second term represents the surface term of the EH action which we call as the Einstein boundary term.
 The Einstein boundary term can be written down as
\begin{eqnarray}\label{5}
\mathcal{A}_{surf}& =&\frac{1}{16\pi G}\int d^{4}x\mathcal{L}_{surf}\nonumber\\ &=&\frac{1}{16\pi G}\int d^{4}x~ 2\partial_{c}\big[\sqrt{-g}~M_{a}{^b}{^c}{^d}\Gamma^{a}_{bd}\big]
\equiv\frac{1}{16\pi G}\int d^{4}x~ \partial_{c}\big[\sqrt{-g}S^c\big]
\end{eqnarray}
where $S^{c}$ is given as follows
\begin{equation}\label{S}
S^{c}=2\sqrt{-g}M_{a}{^b}{^c}{^d}\Gamma^{a}_{bd}~.
\end{equation}
Our aim in this paper is to show that there are examples  where only the surface term is important in the computation of entropy of extremal black holes using the entropy function approch. This is in line with the new observation made in \cite{Majhi:2015pra} where it was argued that the entropy function of the extremal near horizon black holes can be constructed only from the GHY boundary term. We shall demonstrate here explicitly that this is not the case for asymptotically AdS black holes and \textbf{extremal rotating black holes} where the bulk term is also important in the construction of the entropy function.\\ 

\section{Near horizon geometry of  extremal black hole spacetime}\label{cc}
In this section, we briefly discuss the near horizon geometry of extremal black holes . For the sake of simplicity, we consider the spherically symmetric black holes characterized by the following metric in $3+1-$ dimensions
\begin{equation}\label{9}
ds^{2}=-f(\rho)d\tau^{2}+\frac{d\rho^{2}}{f(\rho)}+\rho^{2}d\Omega^{2}~.
\end{equation}
The event horizon is given by the relation  $f(a)=0$. 
  Performing a Taylor expansion of the lapse function around the event horizon \cite{paddy:2010}, and keeping terms quadratic in $(\rho-a)$, we have
\begin{equation}
f(\rho)\approx(\rho-a)f^{\prime}(a)+ \frac{(\rho-a)^{2}}{2}f''(a)~
\end{equation}
 define
  $\kappa=\frac{f^{\prime}(a)}{2}$ is to be the surface gravity.\\ 
 In case of extermal black holes
  $f^{\prime}(a)=0$ since surface gravity $\kappa$ vanishes. 
  Thus, the general form of the near horizon geometry corresponding to a spherically extremal black hole can be written down as
 \begin{eqnarray}\label{12}
 ds^{2} \approx-\frac{(\rho-a)^{2}}{2}f''(a) d\tau^{2}+\frac{dr^{2}}{\frac{(\rho-a)^{2}}{2}f''(a)}+\rho^{2}d\Omega^{2}~.
 \end{eqnarray}
 
 \subsection{Near horizon geometry of extremal Reissner-Nordstr\"om black hole }
 The first solution which we consider is the  Reissner-Nordstr\"om black hole geometry in $3+1$-dimensions. The  metric for this reads \cite{Sen:2007qy}
 \begin{equation}
 ds^{2}=-(1-\frac{a}{\rho})(1-\frac{b}{\rho})d\tau^{2}+\frac{d\rho^{2}}{(1-\frac{a}{\rho})(1-\frac{b}{\rho})}+\rho^{2}(d\theta^{2}+\sin^{2}\theta d\phi^{2})~.
 \end{equation}
 The non-vanishing components of the Faraday tensor are 
 \begin{equation}
 F_{\rho\tau}=\frac{q}{4\pi\rho^{2}};~~
 F_{\theta\phi}=\frac{p}{4\pi}sin\theta~
 \end{equation}
 where $\rho,\theta,\phi,$ and $\tau$ are the spacetime coordinates. The constants $a$ and $b$ are given by 
 \begin{eqnarray}
 a+b=2GM;~~ ab=\frac{G}{4\pi}(q^{2}+p^{2})~
 \end{eqnarray}
 $q, p$ and $M$ represents the electric charge, magnetic charge and the mass of the black hole respectively. The extremal limit ($T_H=0$)  gives the relationship between the mass and charge of the black hole to be \cite{Sen:2007qy}
 \begin{equation}
 M^{2}=\frac{1}{4\pi G}(q^{2}+p^{2})~.
 \end{equation}
 This in turn leads to the following condition
 \begin{equation}
 a=b=\sqrt{\frac{G}{4\pi}(q^{2}+p^{2})}~.
 \end{equation}
The near horizon geometry corresponding to the extremal Reissner-Nordstr\"om black hole then reads
 \begin{eqnarray}\label{11}
 ds^{2}=a^{2}(-r^{2}dt^{2}+\frac{dr^{2}}{r^{2}})+a^{2}(d\theta^{2}+\sin^{2}\theta d\phi^{2});~~
 F_{rt}=\frac{q}{4\pi};~~ F_{\theta\phi}=\frac{p}{4\pi}\sin\theta\nonumber~.
 \end{eqnarray}
 We can observe that the near horizon geometry of the extremal Reissner-Nordstr\"om black hole is $AdS_{2}\times S^{2}$, that is it has $SO(2,1)\times SO(3)$ isometry \cite{Sen:2007qy}.
 The above mentioned black hole spacetime is associated with the following entropy
 \begin{equation}\label{d}
 S_{BH}=\frac{(q^{2}+p^{2})}{4}~.
 \end{equation}

 \subsection{Near horizon geometry of extremal charged BTZ black hole}
We now consider the charged BTZ black hole solution. BTZ black hole is the solution of Einstein field equations in $2+1$-dimensions with negative cosmological constant. The spacetime interval in this theory reads \cite{Martinez:1999qi,Cadoni:2008mw}
\begin{equation}
 ds^{2}=\bigg(-8GM+\frac{\rho^{2}}{l^{2}}-\frac{2Gq^{2}}{\pi}\ln\left(\frac{\rho}{l}\right)\bigg)d\tau^{2}+\frac{d\rho^{2}}{\bigg(-8GM+\frac{\rho^{2}}{l^{2}}-\frac{2Gq^{2}}{\pi}\ln\left(\frac{\rho}{l}\right)\bigg)}+\rho^{2}d\theta^{2}~.
\end{equation}
and the non-vanishing component of Faraday tensor read
\begin{equation}
F_{\tau\rho}=\frac{q}{2\pi \rho}~.
\end{equation}
The extremal condition ($T_H=0$) implies that there is a relationship between the mass and charge of the black hole. This is given as follows
 \begin{equation}
 M=\frac{q^{2}}{8\pi}\bigg(1-2\ln\left(q\sqrt{\frac{G}{\pi}}\right)\bigg)~.
 \end{equation}
 In the extremal case the lapse function reads
 \begin{equation}
  f(\rho)=-Gq^{2}+\frac{2Gq^{2}}{\pi}\ln\bigg(q\sqrt{\frac{G}{\pi}}\bigg)+\frac{\rho^{2}}{l^{2}}-\frac{2Gq^{2}}{\pi}\ln\bigg(\frac{\rho}{l}\bigg)~. 
  \end{equation}
  The near horizon geometry of the extremal charged BTZ black hole, therefore reads
 \begin{equation}\label{BTZ}
 ds^{2}=\frac{l^{2}}{2}\bigg(-r^{2}dt^{2}+\frac{dr^{2}}{r^{2}}\bigg)+r_{h}^{2}d\theta^{2}~
 \end{equation}
 where $r_{h}=ql\sqrt{\frac{G}{\pi}}$ is the radius of the horizon.\\ The non-vanishing component of the Faraday tensor is 
 \begin{equation}
 F_{tr}=\frac{l^{2}q}{2r_{h}}~.
 \end{equation}
 The entropy of the extremal charged BTZ black hole is obtained to be
 \begin{eqnarray}
 S_{BH}=\frac{\pi r_{h}}{2G}=\frac{ql}{2}\sqrt{\frac{\pi}{G}}\label{e}~.
 \end{eqnarray} 
 We observe the near horizon geometry of the  extremal charged BTZ black hole takes the form $AdS_{2}\times S^{1}$ (or has $SO(2,1)\times SO(2)$ isometry). In general one can say that the near horizon geometry of an extremal spherically symmetric black hole has $SO(2,1)\times SO(D-2)$ isometry in $D$- spacetime dimensions \cite{Sen:2007qy}. 
 
   
\section{\textbf{Entropy function of an extremal near horizon black hole from the Einstein Boundary term}}\label{aa}
Entropy function formalism is a powerful way to compute the near horizon geometry and entropy of an extremal black hole when the full black hole geometry is not known. In this formalism, one starts with a function $F$ of the following form \cite{Sen:2005wa,Sen:2007qy}
\begin{eqnarray}\label{F}
F(v,e,p)=\frac{1}{16\pi G}\int d\theta d\phi\sqrt{-g}\big[R-2\Lambda\big]+\int d\theta d\phi \sqrt{-g}\mathcal{L}_{m}~.
\end{eqnarray}
 Using the identity 
 \begin{equation}
 \sqrt{-g}R=\sqrt{-g}\mathcal{L}_{quad}+\mathcal{L}_{surf}
 \end{equation}
  leading to eq.(\ref{2}), eq.(\ref{F}) can be recast as  
\begin{eqnarray}
F(v,e,p)=\frac{1}{16\pi G}\int d\theta d\phi\big[\sqrt{-g}(\mathcal{L}_{quad}-2\Lambda)+\mathcal{L}_{surf}\big]+\int d\theta d\phi \sqrt{-g}\mathcal{L}_{m} ~.
\end{eqnarray}
We shall show that for $\Lambda=0$,
  $\mathcal{L}_{quad}$ when computed using the near horizon metric for the extremal black hole yields a vanishing contribution and the total geometrical information comes from the surface term $\mathcal{L}_{surf}$. However, in case of gravity solutions with non-vanishing cosmological constant ($\Lambda\neq 0$) this is not true. In this case we do get a contribution from the cosmological term in the bulk.
\subsection{Extremal Reissner-Nordstr\"om black hole}
The metric ansatz for the near horizon spherically symmetric extremal black hole in $3+1$- dimensions reads \cite{Sen:2007qy}
\begin{equation}\label{RN}
ds^{2}=v_{1}(-r^{2}dt^{2}+\frac{dr^{2}}{r^{2}})+v_{2}(d\theta^{2}+\sin{\theta}d\phi^{2})~.
\end{equation}
The information about the matter part gets specified by the following components of the Faraday tensor 
\begin{equation}\label{13}
F_{rt}=e;~ F_{\theta\phi}=\frac{p}{4\pi}\sin \theta~
\end{equation}\\
where $v_{1},v_{2},e,p$ are constants.
  The non-zero Christofell symbols for the metric given in eq.(\ref{RN}) reads
\begin{eqnarray}\label{14}
\Gamma^{t}{_r}{_t}=\frac{1}{r},~\Gamma^{r}{_t}{_t}=r^{3},~ \Gamma^{r}{_r}{_r}=-\frac{1}{r},~ \Gamma^{\theta}{_\phi}{_\phi}=-\sin\theta \cos\theta,~
\Gamma^{\phi}{_\theta}{_\phi}=\cot\theta~.
\end{eqnarray}
 Using the above Christofell sysmbols, we can obtain the non-vanishing components of $S^{c}$ (given in eq.(\ref{S})). This reads 
\begin{eqnarray}\label{15}
S^{r}=-\frac{2r}{v_{1}};~ S^{\theta}=-\frac{2\cot\theta}{v_{2}}~.
\end{eqnarray}
We now define a function $F(v_{1},v_{2},e,p)$ by integrating the total Lagrangian density on the transverse coordinates corresponding to the extremal near horizon geometry of the black hole \cite{Sen:2005wa,Sen:2007qy}. This reads 
\begin{eqnarray}\label{16}
F(v_{1},v_{2},e,p)&=&\frac{1}{16\pi G}\int d\theta d\phi\big[\sqrt{-g}\mathcal{L}_{quad}+\mathcal{L}_{surf}\big]+\int d\theta d\phi \sqrt{-g}\mathcal{L}_{m}\nonumber\\
&=&\frac{1}{16\pi G}\int d\theta d\phi \mathcal{L}_{surf}+\int d\theta d\phi \sqrt{-g}\mathcal{L}_{m}~
\end{eqnarray}
since $\mathcal{L}_{quad}$ vanishes for the near horizon geometry. Computing $\mathcal{L}_{sur}$ and $\mathcal{L}_{m}$ on the near horizon geometry, we get
\begin{eqnarray}
\mathcal{L}_{surf}&=&2(v_{1}-v_{2})\sin\theta \label{ls}\\
\sqrt{-g}\mathcal{L}_{m}&=&v_{1}v_{2}\sin\theta \bigg(\frac{e^{2}}{2v_{1}^{2}}-\frac{1}{2v_{2}^2}\bigg(\frac{p}{4\pi}\bigg)^{2}\bigg)\label{lm}~.
\end{eqnarray}
This leads to the following form for $F(v_{1},v_{2},e,p)$: 
	 
\begin{eqnarray}\label{28}
F(v_{1},v_{2},e,p)&
=& \frac{1}{2G}(v_{1}-v_{2})+2\pi\bigg[e^{2}\frac{v_{2}}{v_{1}}-\bigg(\frac{p}{4\pi}\bigg)^{2}\frac{v_{1}}{v_{2}}\bigg]~.
\end{eqnarray}
With the above  result for $F(v_{1},v_{2},e,p)$ in hand, the entropy function $E(v_{1},v_{2},e,q,p)$ is defined as
\begin{eqnarray}\label{29}
E(v_{1},v_{2},e,p,q)&=&2\pi \bigg[qe-F(v_{1},v_{2},e,p)
\bigg]\nonumber\\&=&2\pi \bigg[qe-\frac{1}{2G}(v_{1}-v_{2})-2\pi\bigg\{e^{2}\bigg(\frac{v_{2}}{v_{1}}\bigg)-\bigg(\frac{p}{4\pi}\bigg)^{2}\frac{v_{1}}{v_{2}}\bigg\}\bigg]~.
\end{eqnarray}
The extremization conditions for $E(v_{1},v_{2},e,p)$ are given below \cite{Sen:2007qy}
\begin{eqnarray}
\frac{\partial E}{\partial e}=0\label{a};~~
\frac{\partial E}{\partial v_{1}}=0\label{b};~~
\frac{\partial E}{\partial v_{2}}=0 \label{c}~.
\end{eqnarray}
Extremization of the entropy function ($E$) with respect to $e$ gives
\begin{eqnarray}
e=\frac{q}{4\pi}\frac{v_{1}}{v_{2}}~.
\end{eqnarray}
We now substitute the above relation in eq.(\ref{29}) and by extremising eq.(\ref{29}) with respect to $v_{1}$ and $v_{2}$ we obtain 
\begin{equation}\label{35}
v_{1}=v_{2}=\frac{G}{4\pi}(q^{2}+p^{2})~.
\end{equation}
By substituting the obtained values of $v_1$ and $v_2$ in eq.(\ref{29}), we obtain the extremum value of the entropy function to be  
\begin{equation}\label{36}
E\mid_{extremum}=\frac{1}{4}(q^{2}+p^{2})~.
\end{equation}
The above  expression matches exactly with that of $S_{BH}$ (given in eq.(\ref{d})). This in turn means that
\begin{eqnarray}
E\mid_{extremum} = S_{BH}~.
\end{eqnarray}
In this example we have thus shown that the entropy function can be derived by considering only the surface term of the Einstein-Hilbert action which we call as the Einstein boundary term. It is therefore seems as if the boundary term in the Einstein-Hilbert contribution solely in the construction of the entropy function. However, as we will see in the next example , this is not true.

\subsection{Charged extremal BTZ black hole}
We now consider the extremal charged BTZ black hole, which is a spherically symmetric charged AdS black hole in $2+1$-dimensions. The motivation to discuss this theory lies in the fact that in this case we shall see that the Einstein boundary term is not enough to calculate the entropy of the black hole as there is a non-vanishing contribution from the cosmological constant term. The full action of the theory reads \cite{Martinez:1999qi,Cadoni:2008mw}
\begin{equation}\label{37}
\mathcal{A}=\frac{1}{16\pi G}\int d^{3}x\sqrt{-g}\bigg[(R+\frac{2}{l^{2}})-4\pi G F_{ab}F^{ab}\bigg]~;~\Lambda=-\frac{1}{l^{2}} ~
\end{equation}
where $R$ is the Ricci scalar, $l$ is the AdS radius and $F_{ab}$ is the Faraday tensor. We now take the following $AdS_2\times S^1$ ansatz for the near horizon geometry
\begin{eqnarray}
ds^{2}=v_{1}\bigg(-r^{2}dt^{2}+\frac{dr^{2}}{r^{2}}\bigg)+v_{2}^{2}d\theta^{2}\label{38}~.
\end{eqnarray}
The non-vanishing components of the Faraday tensor reads $F_{tr}=e\label{39}$. The parameters $v_{1}$,$v_{2}$ characterize the near horizon geometry and $e$ is the electric field. Now we  define a function $F(v_{1},v_{2},e)$ in the following way
\begin{equation}\label{40}
F(v_{1},v_{2},e)=\frac{1}{16\pi G}\int d\theta \big[\sqrt{-g}(\mathcal{L}_{quad}+\frac{2}{l^{2}})+\mathcal{L}_{surf}-4\pi G\sqrt{-g}F_{ab}F^{ab}\big]~.
\end{equation} 
 Coumputation of $\mathcal{L}_{quad}$ for the near horizon geometry (given in eq.(\ref{38})) yields a vanishing contribution while from the rest of the terms we obtain the following
\begin{equation}\label{41}
F(v_{1},v_{2},e)=-\frac{v_{2}}{4G}+\frac{v_{1}v_{2}}{4Gl^{2}}+\pi e^{2}\frac{v_{2}}{v_{1}}~
\end{equation}
with $\mathcal{L}_{surf}$ and $\mathcal{L}_{m}$ following form
\begin{eqnarray}
\mathcal{L}_{surf}&=&-2v_{2}\label{lsbtz}\\
\sqrt{-g}\mathcal{L}_{m}&=&-e^{2}\frac{v_{2}}{2v_{1}}
\end{eqnarray}
 The entropy function $E(v_{1},v_{2},e,q)$ therefore reads 
\begin{eqnarray}\label{EE}
E(v_{1},v_{2},e,q)&=&2\pi\bigg(eq-F(v_{1},v_{2},e)\bigg)\nonumber\\
&=&2\pi\bigg(eq+\frac{v_{2}}{4G}-\frac{v_{1}v_{2}}{4Gl^{2}}-\pi e^{2}\frac{v_{2}}{v_{1}}\bigg)\label{43}~.
\end{eqnarray}
Similar to the previous example, we consider the extremization of $E(v_{1},v_{2},e,q)$ with respect to $e$. This leads to 
\begin{equation}\label{44}
e=\frac{q}{2\pi}\frac{v_{1}}{v_{2}}~.
\end{equation}
Substituting this expression for $e$ into eq.(\ref{43}) and by extremizing $E(v_{1},v_{2},e,q)$ with respect to the parameter $v_{1}$ we get
\begin{equation}\label{45}
v_{2}=ql\sqrt{\frac{G}{\pi}}~.
\end{equation}
Similarly, extremization with respect to $v_{2}$ yields 
\begin{equation}\label{46}
v_{1}=\frac{l^{2}}{2}~.
\end{equation}
 Substituting these obtained values of $v_1$, $v_2$ and $e$ in eq.(\ref{EE}), we get the extremum value for the entropy function. This reads
\begin{equation}\label{47}
E\mid_{extremum}=\frac{ql}{2}\sqrt{\frac{\pi}{G}}=S_{BH}~.
\end{equation}
 We again observe that this extremum value of $E$ matches exactly with the expression for $S_{BH}$ (given in eq.(\ref{e})). We would like to stress that in this case the surface term (Einstein boundary term) alone does not lead to the correct entropy function which leads to the entropy on extremisation. We need to include the contribution coming from the cosmological constant in the bulk action. Hence it is not true that the boundary term alone can lead to the correct entropy funtion as mentioned in \cite{Majhi:2015pra}.\\
  We now show that the GHY boundary term computed with the near horizon metric is identical with the Einstein boundary term and hence we can start with a modified function $F(v_{1},v_{2},e)$ to get the entropy function $E(v_{1},v_{2},e,q)$.
  The Lagrangian density corresponding to GHY term reads \cite{Paddy}
 \begin{equation}\label{48}
 \mathcal{L}_{GHY}=\frac{1}{8\pi G}\partial_{c}(\sqrt{-g}K_{(i)}N^{c}_{(i)})~.
 \end{equation}
 where $K_{(i)}$ is the trace of the extrinsic curveture. This quantity $K_{(i)}$ is usually defined in terms of the normal vector ($N^{a}$) as 
 \begin{equation}\label{49}
 K_{(i)}=-\nabla_{a}N^{a}_{(i)}=-\frac{1}{\sqrt{-g}}\partial_{a}(\sqrt{-g}N^{a}_{(i)})~.
 \end{equation}
 where \textquotedblleft$i$\textquotedblright  in the subscript representing which kind of surface one is working with. For example if $i=t$ then the normal is taken on the $t=constant$ surface which is a space-like surface. We define a function $F(v_{1},v_{2},e)$ as
 \begin{equation}\label{50}
  F(v_{1},v_{2},e)=\int d\theta \sqrt{-g}\bigg(\mathcal{L}_{GHY}+\frac{1}{16 \pi G}\frac{2}{l^{2}}\bigg)-\frac{1}{4}\int d\theta \sqrt{-g}F_{ab}F^{ab}~.
 \end{equation} 
 Now we will calculate the $\sqrt{-g}\mathcal{L}_{GHY}$ and show that it is identical to the Einstein boundary term. 
 \begin{eqnarray}{\label{GHY}}
 \sqrt{-g}\mathcal{L}_{GHY}&=&\frac{1}{8\pi G}\partial_{a}(\sqrt{-g}K_{(i)}N^{a}_{(i)})\nonumber\\
 &=&\frac{1}{8\pi G}\bigg[\partial_{a}(\sqrt{-g}K_{(t)}N^{a}_{(t)})+\partial_{a}(\sqrt{-g}K_{(r)}N^{a}_{(r)})+\partial_{a}(\sqrt{-g}K_{(\theta)}N^{a}_{(\theta)})\bigg]~.
 \end{eqnarray}
 $N^{a}_{(t)}=\bigg(\frac{1}{\sqrt{rv_{1}}},0,0\bigg)$ is the normal vector on $t=constant$ surface (which is a spacelike surface),
  $N^{a}_{(r)}=\bigg(0,\frac{r}{v_{1}},0\bigg)$ is the unit normal on the $r=constant$ surface (which is a timelike surface),
   and $N^{a}_{(\theta)}=\bigg(0,0,\frac{1}{v_{2}}\bigg)$ is unit normal on $\theta=constant$ surface.
   With all these normal vectors, $K_{(i)}$ is obtained to be 
   \begin{equation}{\label{K}}
   K_{(t)}=0~,~K_{(r)}=-\frac{1}{\sqrt{v_{1}}}~,~K_{(\theta)}=0 ~.
   \end{equation}
   This then leads to
   \begin{equation}{\label{ghy}}
   \sqrt{-g}\mathcal{L}_{GHY}=-\frac{v_{2}}{8\pi G}~
   \end{equation}
   which is identical to the Einstein boundary term as discussed earlier eq.(\ref{lsbtz}). And rest of the calculation to compute the entropy via entropy function formalism follows as discussed eariler for charged extremal black hole.
 
 
  \subsection{Extremal Kerr Black hole} 
  
Here we show that for extremal rotating Kerr black hole, a non-vanishing contribution comes from the bulk term. This feature is present also for the Kerr-Newman solution. The near horizon geometry of the extremal Kerr black hole reads \cite{a.sen}
  \begin{eqnarray}
  ds^{2}&=&\Omega^{3}(\theta)e^{2\psi(\theta)}\bigg(-r^{2}dt^{2}+\frac{dr^{2}}{r^{2}}+\beta^{2}d\theta^{2}\bigg)+e^{-2\psi(\theta)}\bigg(d\phi-\alpha rdt\bigg)^{2}~.\label{kerrm}
  \end{eqnarray} 
For the above metric $\mathcal{L}_{quad}$ yields a non-vanishing contribution. In this case we have
\begin{eqnarray}
\sqrt{-g}\mathcal{L}_{quad}&=&\frac{\alpha^{2}\beta e^{-4\psi(\theta)}}{2\Omega(\theta)}-\frac{2\Omega(\theta)\psi^{\prime}(\theta)}{\beta}+\frac{2{\Omega^{\prime}}^{2}(\theta)}{\beta\Omega(\theta)}\label{lqk}\\
\mathcal{L}_{surf}&=&-\bigg[\Omega(\theta)\beta+\frac{1}{\beta}\bigg(2\frac{d^{2}\Omega}{d\theta^{2}}+\frac{d\Omega}{d\theta}\frac{d\psi}{d\theta}+\Omega\frac{d^{2}\psi}{d\theta^{2}}\bigg)\bigg]\label{kerrb}~.
\end{eqnarray}
Hence the function $F(\Omega(\theta),\psi(\theta),\alpha,\beta)$ takes the form
\begin{eqnarray}\label{kF}
F(\Omega(\theta),\psi(\theta),\alpha,\beta)&=&\frac{1}{8G}\int d\theta \frac{\alpha^{2}\beta^{2}e^{-4\psi(\theta)}-4\Omega^{2}(\theta)(\beta^{2}+{\psi^{\prime}}^{2}(\theta))+4{\Omega^{\prime}}^{2}(\theta)}{2\Omega(\theta)\beta}\nonumber
\\&-&\frac{1}{4G}\int d\theta \frac{1}{\beta}\bigg(2\frac{d^{2}\Omega}{d\theta^{2}}+\frac{d\Omega}{d\theta}\frac{d\psi}{d\theta}+\Omega\frac{d^{2}\psi}{d\theta^{2}}\bigg)~.
\end{eqnarray}
 This is the form that one gets from the usual definition of the function $F(\Omega(\theta),\psi(\theta),\alpha,\beta)$ \cite{a.sen}. Now one can follow the standard procedure to find the entropy from the entropy function formalism.



\section{Conclusion}
In this paper we show that the entropy function corresponding to a static extremal (asymptotically flat) black hole can be formulated entirely from the Einstein boundary term along with the matter term. This observation is a consequence of the fact that the value of the bulk term $\mathcal{L}_{quad}$ is zero for the near horizon geometry. This in turn means that the term $\mathcal{L}_{surf}$ contains all the geometric information needed to construct the entropy function.  
 However, this statement is only valid for spacetimes with vanishing cosmological constant. For asymptotically $AdS$ spacetimes, the bulk term reads $\left(\mathcal{L}_{quad}-2\Lambda\right)$. This in turn means that the entropy function has a non-vanishing contribution from the bulk term due to the presence of the cosmological constant. Further, the story gets more interesting when we consider an extremal rotating black hole where the near horizon geometry has the form $AdS_2 \times U(1)$. In this case although the EH action does not contain a cosmological constant but  a finite, non-vanishing contribution from the bulk term $\mathcal{L}_{quad}$ exists. These observations lead to the fact that one can define a general function as
\begin{eqnarray}
	F(\vec{v},e,p)=\frac{1}{16\pi G}\int d\theta d\phi\big[\sqrt{-g}(\mathcal{L}_{quad}-2\Lambda)+\mathcal{L}_{surf}\big]+\int d\theta d\phi \sqrt{-g}\mathcal{L}_{m} ~.
\end{eqnarray}
We conclude by making another interesting observation that, the Ricci scalar for the full Reissner-Nordstr\"om
solution is nonzero, but the Ricci scalar corresponding to the extremal near horizon geometry vanishes.
However, for the full Kerr solution the Ricci scalar vanishes as it is a vaccum solution, but for the extremal
near horizon Kerr geometry it does not vanish. There may be a deeper reason behind this observation and this
surely needs further investigation in future.
\section*{Acknowledgement}
AS would like to acknowledge the support by CSIR, Govt. of India for a Senior Research Fellowship. ARC acknowledges the support by the SNBNCBS for a Junior Research Fellowship.

\end{document}